\documentclass[aip,amsmath,amssymb,reprint,]{revtex4-1}

\usepackage{graphicx}
\usepackage{dcolumn}
\usepackage{bm}
\usepackage[utf8]{inputenc}
\usepackage[T1]{fontenc}
\usepackage{mathptmx}
\usepackage{xcolor}
\usepackage{siunitx}

\usepackage{hyperref}

\begin{document}

\preprint{C_Suergers_JAP2020}

\title[Anomalous Nernst effect in ferromagnetic Mn$_5$Ge$_3$C$_{x}$ films]{Anomalous Nernst effect in ferromagnetic Mn$_5$Ge$_3$C$_{x}$ thin films \\on insulating sapphire}

\author{R. Kraft}
\email{rainer.kraft@kit.edu}
\author{S. Srichandan}
\author{G. Fischer}
\author{C. S\"urgers}
\email{christoph.suergers@kit.edu}
\affiliation{Physikalisches Institut, Karlsruhe Institute of Technology, Wolfgang Gaede Str.~1, 76131 Karlsruhe, Germany}

\date{\today}

\begin{abstract}
	Investigating the thermoelectric properties of ferromagnets is important for the development of future microelectronic devices for efficient energy conversion purposes. Ferromagnetic Mn$_5$Ge$_3$C$_{x}$ thin films with a Curie temperature up to $T_{\rm \rm C}=450\,$K well above room temperature are potential candidates for spintronic applications by integration into CMOS heterostructures. In this work, the thermoelectric power, in particular, the anomalous Nernst effect (ANE) has been investigated experimentally for magnetron sputtered thin films on sapphire (11\=20) substrates. The ANE gradually increases with increasing carbon content $x$ up to a maximum value obtained for $x = 0.8$ in line with earlier investigations of the magnetization and anomalous Hall effect. The ANE is strongly enhanced by a factor three compared to the parent Mn$_5$Ge$_3$ compound. However, for $x = 0.8$ we observe a clear deviation of the calculated ANE from the measured values. 
\end{abstract}

\maketitle

\section{Introduction}

Thermal transport in ferromagnetic metals interlaced with spin phenomena can give rise to several interesting effects like the spin dependent Seebeck and Peltier effects \cite{Slachter, Flipse}, magnon drag thermopower \cite{Watzman, Srichandan}, and the thermal spin transfer torque \cite{Hatami}. Among these effects the anomalous Nernst effect (ANE), which refers to the generation of a transverse voltage in the presence of a heat current and spontaneous magnetization, has recently received a lot of attention \cite{Mizuguchi2019}. This is not only due to the potential it holds for increased conversion efficiency in thermoelectric devices but also because this effect has been observed in a wide range of materials such as chiral antiferromagnets \cite{Ikhlas,Guo,Hanasaki2008}, Weyl semimetals \cite{Liang, Chernodub,Wuttke2019}, and group-IV dichalcogenides \cite{Yu}. The focus is not limited to the mere observation of this effect in novel materials but also to devise new methods of enhancing the anomalous Nernst coefficient, in particular in magnetic conductors \cite{Chuang,Kannan,Uchida2}.

From a practical point of view, it is not the conventional metals but rather alloys, often magnetic, which exhibit a high thermoelectric conversion efficiency. Additionally, a high Curie temperature $T_{\rm C}$ is a desirable property among ferromagnets which can facilitate the integration into current electronic industry. Therefore we have chosen to study the thermoelectric effects in Mn$_5$Ge$_3$C$_{x}$ thin films grown on sapphire substrates. Ferromagnetic Mn$_5$Ge$_3$C$_{x}$ ($0 \le x \le 1$) is of particular interest for future spintronic devices due to its compatibility with CMOS device technology \cite{Thanh,Fischer}. It has an uniaxial magnetic anisotropy along the c axis of the hexagonal D8$_8$ structure belonging to space group \textit{P}$_6$/mcm and the Curie temperature of Mn$_5$Ge$_3$, $T_{\rm C}=300\,$K, can be boosted up to $450\,$K by incorporation of carbon into the crystalline lattice of Mn$_5$Ge$_3$C$_{x}$ \cite{Gajdzik,Surgers}. The structural, magnetic, and transport properties of this particular material have been previously investigated in detail \cite{Thanh,Gajdzik,Surgers}. 

The origin of the ANE in various metals including itinerant ferromagnets, oxides, and chalcogenides is often discussed in terms of an unified theory involving the emergence of the anomalous Hall effect (AHE) \cite{Asamitsu}. The scattering mechanism, be it intrinsic or extrinsic, directly links the AHE and ANE coefficients. The anisotropic magnetoresistance (AMR) and anomalous Hall effect of ferromagnetic Mn$_5$Ge$_3$C$_{x}$ and isostructural Mn$_5$Si$_3$C$_{0.8}$ thin films have been investigated previously \cite{Surgers}.

Here we report on measurements of the ANE as a function of temperature for Mn$_5$Ge$_3$C$_{x}$ films with different carbon concentrations $x$ and for Mn$_5$Si$_3$C$_{0.8}$ films deposited on single crystalline sapphire substrates. Because the samples have a magnetization oriented almost in the plane of the film we have chosen to study the perpendicular ANE where the magnetic field $H$ is applied in plane, the temperature gradient is generated out of plane, and the ANE thermovoltage is measured perpendicular to both, field and temperature gradient. Furthermore, this configuration bears the advantage that unintentional temperature gradients are minimized which are often observed in ANE thin-film experiments with in-plane heat transport where the thermal conduction through the 10$^4$-order thicker substrate overwhelms the thermal transport \cite{Huang}.

\section{Experimental}

28-nm thick Mn$_5$Ge$_3$C$_{x}$ films were deposited on (11\=20)-oriented Al$_2$O$_3$ substrates by magnetron sputtering at a substrate temperature of \SI{400}{\celsius} and a base pressure of $10^{-7}\,$mbar, as described earlier \cite{Surgers}. For comparison, a film of ferromagnetic isostructural 28-nm thick Mn$_5$Si$_3$C$_{0.8}$ was deposited at \SI{450}{\celsius}. The films grow in a polycrystalline fashion as checked by x-ray diffraction (not shown) \cite{Gajdzik}. Therefore, the magnetization is preferentially oriented in the plane of the film as opposed to (0001)-oriented films grown on Ge(111) which show a tendency to out-of-plane magnetization for thicknesses larger than $\approx 20\,$nm \cite{Spiesser,Michez}.

The samples were magnetically characterized by SQUID magnetometry and the resistivity and Hall effect were acquired in a physical property measurement system (PPMS). For thermoelectric measurements, the substrate was glued to the copper holder of the PPMS with an external nanovoltmeter connected by 50-\SI{}{\micro\meter} thick Pt wires to the film to measure the thermovoltage. A \SI{2}{\kilo\ohm} resistive heater was used to heat a copper foil placed on top of the film, electrically insulated by a thin Kapton foil, to ensure a homogeneous temperature distribution. The temperature gradient was recorded in a separate run by measuring the temperature at the top and the bottom surface of the chip and applying the same heater currents that were used during the thermovoltage measurements. 

\begin{figure}
	\centering
	\includegraphics[width=\columnwidth]{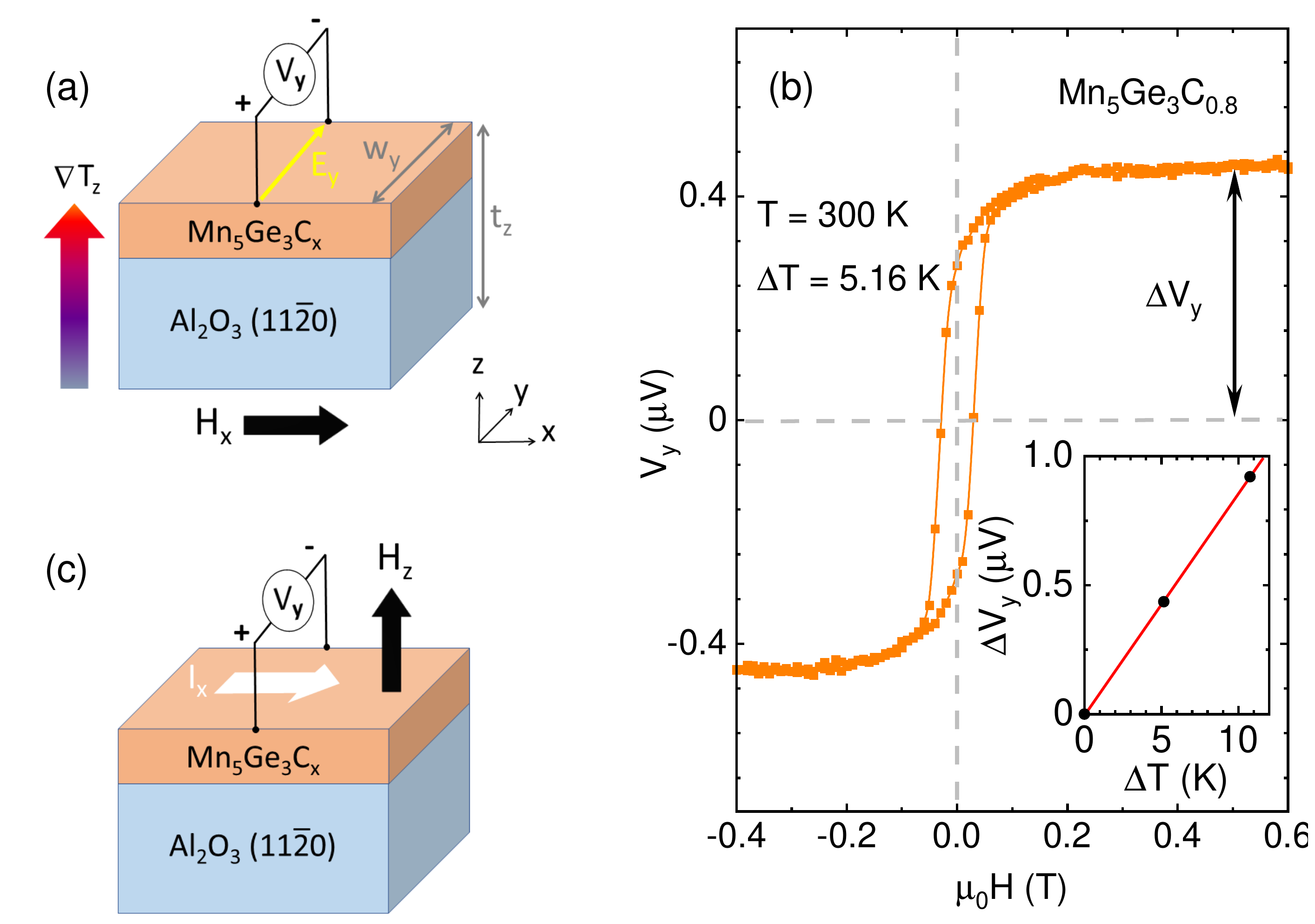}
	\caption{(a) Configuration for measuring the anomalous Nernst effect. (b) Hysteresis of the thermovoltage $V_y$ vs.~applied magnetic field $H$. Inset shows a linear behavior of the thermovoltage $\Delta V_y$ at magnetic saturation vs.~the temperature difference $\Delta T$ between the hot and cold side of the sample. (c) Configuration for measuring the Hall effect.} 
	\label{F1}
\end{figure}

Fig.~\ref{F1}(a) shows the configuration for measuring the ANE. In presence of a magnetization \textbf{\textit{M}} the ANE creates an electric field $\textbf{E}= S_{\rm ANE} {\nabla} T \times \textbf{M}$ where $S_{\rm ANE}$ is the anomalous Nernst coefficient and ${\nabla} T$ is the temperature gradient \cite{Bauer}. The ANE for the present configuration - assuming a temperature gradient only present perpendicular to the plane - is 

\begin{equation}
\label{ANE}
E_y=S_{\rm ANE} \frac{\partial{T}}{\partial{z}} \frac{M_x}{\left|\textbf{\textit{M}}\right|}\,. 
\end{equation} 

Experimentally, the thermoelectric coefficient $S_{yz}(H)= [-V_y(H)/w_y]/[{\Delta T}/t_z]$ was measured, where $V_y$, ${\Delta T}$, $w_y$, and $t_z$ are the voltage, temperature difference between hot and cold surface, and width and thickness of the sample (substrate plus film), respectively. $S_{\rm ANE}=S_{yz}(\Delta V_y)$ corresponds to the thermovoltage $\Delta V_y$ reached at magnetic saturation for which $M_x = \left|\textbf{\textit{M}}\right|$. An exemplary hysteresis curve of the measured thermovoltage at 300 K with indicated $\Delta V_y$ is presented in Fig.~\ref{F1}(b). The inset shows that $\Delta V_y(H)$ increases with increasing $\Delta T$, as expected from Eq.~\ref{ANE}. The Hall resistivity $\rho_{yx}= (V_y/I_x) t_{film}$ was determined in a magnetic field oriented along $z$ while passing a current $I_x$ through the film and measuring the transverse voltage $V_y$, as shown in Fig.~\ref{F1}(c).

\section{Results}
\begin{figure}
	\centering
	\includegraphics[width=\columnwidth]{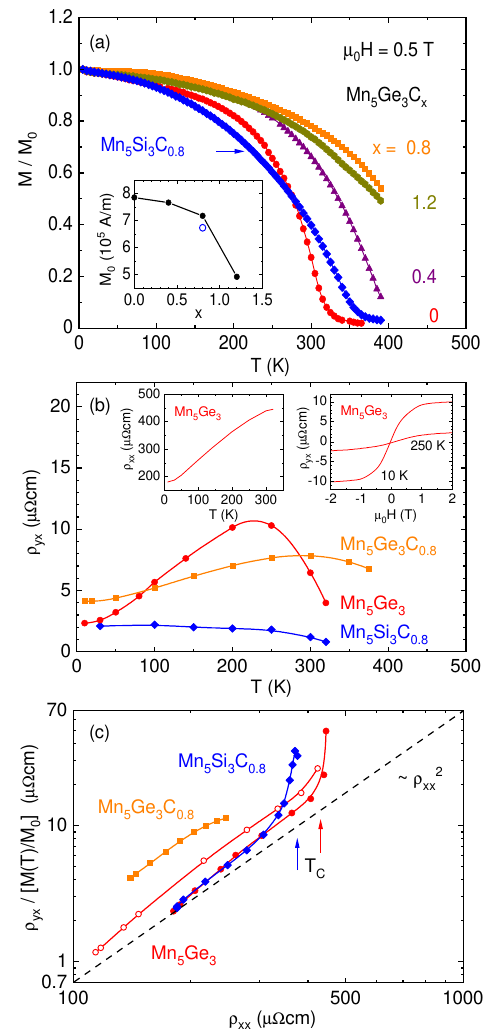}
	\caption{(a) Normalized magnetization $M/M_0$ vs.~temperature $T$ for all investigated films in a magnetic field $\mu_0H=0.5\,$T oriented in the plane of the film. $M_0$ is the saturation magnetization taken from the magnetization curves $M(H)$ at $T=10\,$K, see Fig.~\ref{F3}. The inset shows $M_0$ for all samples where the blue open symbol represents Mn$_5$Si$_3$C$_{0.8}$. (b) Hall resistivity $\rho_{yx}$ vs.~$T$. Insets show the longitudinal resistivity $\rho_{xx}$ vs.~$T$ (left panel) and $\rho_{yx}$ vs.~applied magnetic field $H$ at two temperatures (right panel) for Mn$_5$Ge$_3$. (c) Log-log plot of $\rho_{yx}(T)/[M(T)/M_0]$ vs. $\rho_{xx}(T)$ (closed symbols). Data for a 50-nm thick Mn$_5$Ge$_3$ film from Ref.~\onlinecite{Surgers} are shown as open symbols for comparison. The black dashed line indicates a behavior $\rho_{yx}(T)/[M(T)/M_0] \propto \rho_{xx}^2(T)$.} 
	\label{F2}
\end{figure}

\begin{figure*}
	\includegraphics[width=1.8\columnwidth]{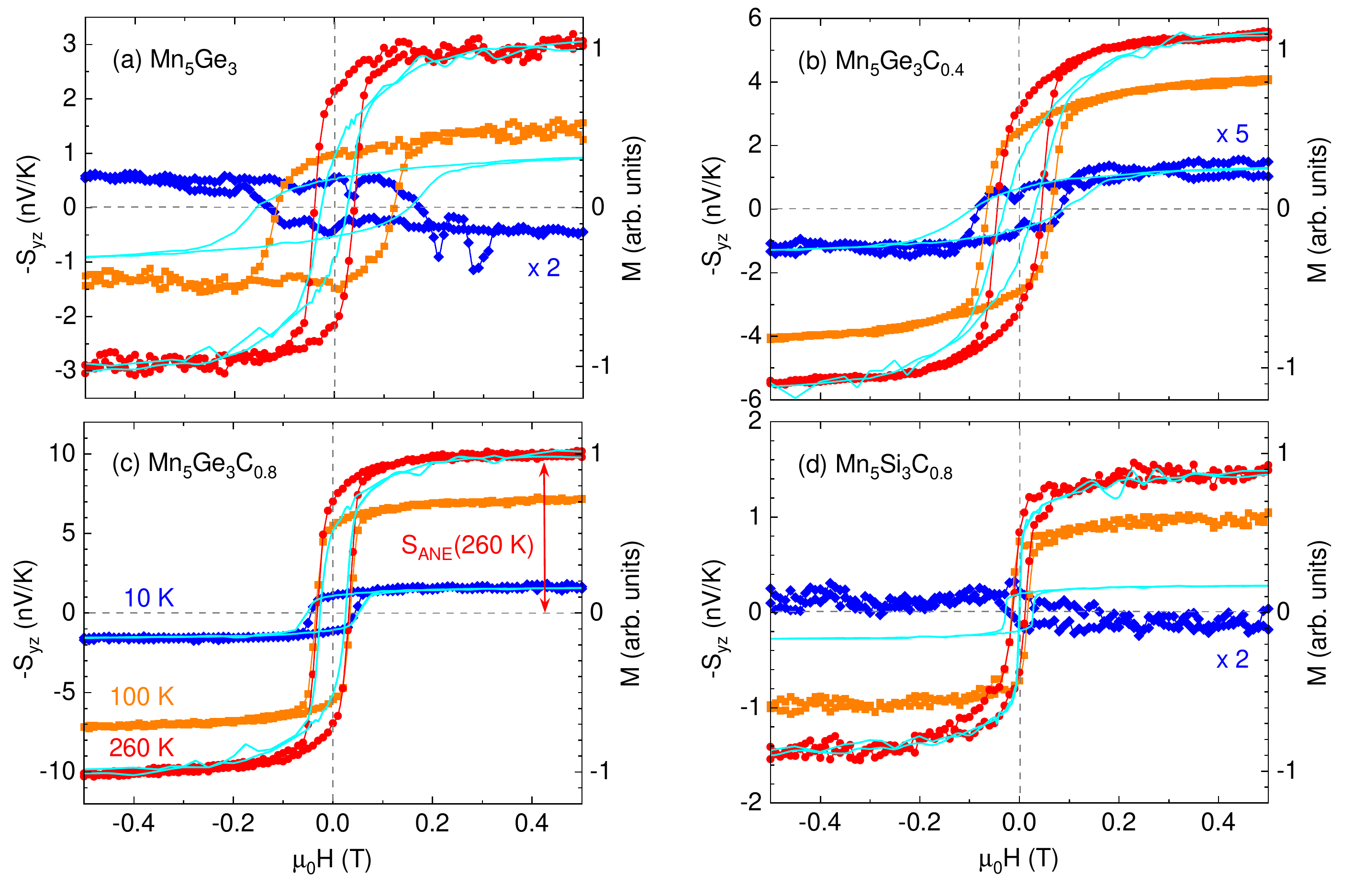}
	\caption{$S_{yz}$ (symbols) and magnetization $M$ (lines in cyan blue) vs.~magnetic field $H$ for (a-c) Mn$_5$Ge$_3$C$_{x}$ with $x=0$, $x=0.4$, and $x=0.8$, respectively, and (d) Mn$_5$Si$_3$C$_{0.8}$ at selected temperatures as indicated in panel (c).} 
	\label{F3}
\end{figure*}

To begin with, we briefly present the magnetic properties of the investigated films. Details of the magnetic properties of Mn$_5$Ge$_3$C$_{x}$ and Mn$_5$Si$_3$C$_{x}$ films on different substrates have been reported earlier \cite{Thanh,Surgers}. Fig.~\ref{F2}(a) shows the magnetization $M$ vs.~temperature $T$ of the 28-nm thick films indicating a gradual increase of the Curie temperature $T_{\rm C}$ with increasing carbon content $x$, from $T_{\rm C}=300\,$K for $x=0$ to $T_{\rm C} \approx 450\,$K for $x=0.8$ \cite{Thanh,Spiesser,Surgers, Michez}. The moderate decrease of the saturation magnetization $M_0$ with increasing $x$ up to $x=0.8$ (see inset) is in agreement with previous results with 800 kA/m corresponding to a magnetic moment of $2\mu_{\rm B}$/Mn for this compound. For $x=1.2$, $M_0$ is strongly reduced. This sample is already in an ``overdoped'' regime where all interstitial sites of the crystalline lattice are filled by carbon and the excess carbon content induces formation of other phases in addition to the major phase with Mn$_5$Ge$_3$ structure \cite{Thanh,Surgers}. 

The films show a metallic behavior of their resistivity $\rho_{xx}(T)$ and an anomalous Hall resistivity $\rho_{yx}(T)$ which decreases with increasing temperature when approaching $T_{\rm C}$, see Fig.~\ref{F2}(b). The dependence of the Hall resistivity on the longitudinal resistivity shows a behavior $\rho_{yx}(T)/M(T) \propto \rho_{xx}^2(T)$ for $T \ll T_{\rm C}$ as indicated by the dashed line in Fig.~\ref{F2}(c) in agreement with earlier data obtained on 50-nm thick films \cite{Surgers}. The strong increase at high $\rho_{xx}$ is simply due to the temperature approaching $T_{\rm C}$. The observed behavior confirms that due to their high resistivity the films are in the intermediate Hall conductivity regime where the extrinsic skew-scattering mechanism does not play a role and the AHE is dominated by intrinsic Berry-phase contribution and the side-jump mechanism \cite{Miyasato,Nagaosa}. 

Next, we investigate the ANE of our films. Fig.~\ref{F3} shows the measured thermoelectric coefficient $S_{yz}(H) \propto V_y$ for a full cycle of the magnetic field together with respective magnetization curve $M(H)$ of each film at selected temperatures. Clear thermoelectric signals of a few nV/K have been recorded with a sign reversal when $H$ is reversed. Magnetic field independent offsets have been subtracted to center the curves around zero voltage. Except for the lowest temperatures, each film shows a hysteresis in $S_{yz}(H)$ and $M(H)$ of similar shape and coercivity, characteristic of the ANE. The strongest signal is observed for $x=0.8$ being of considerable magnitude even at $10\,$K. 

For Mn$_5$Ge$_3$ and Mn$_5$Si$_3$C$_{0.8}$, see Fig.~\ref{F3}(a) and (d), respectively, we observe an inversion of the hysteresis of $S_{yz}(H)$ with respect to $M(H)$ at temperatures below $\approx 40\,$K, \textit{i.e}., a change of the sign of $S_{yz}(H)$ compared to curves obtained at $T > 40\,$K but still with similar coercivity as the magnetic hysteresis. However, note that the thermovoltage at low temperatures is very small and should vanish for $T \rightarrow 0$ according to Nernst's theorem. An additional thermovoltage could occur from the so-called planar Nernst effect (PNE) due to an in-plane component of the temperature gradient \cite{Schmid}. However, this would give rise to a voltage signal which would be an even function with respect to $H$ similar to the AMR, in contrast to what is observed. Furthermore, the magnetization curves do not show an indication of a change of the magnetization behavior between $10\,$K and $40\,$K but the ANE does. A sign change of the \textit{regular} Hall coefficient from negative ($T < 100\,$K) to positive ($T > 100\,$K) has been observed for Mn$_5$Ge$_3$C$_{x}$ but not for Mn$_5$Si$_3$C$_{0.8}$ \cite{Surgers}. We note that while contributions from electrons and holes cancel in the Hall effect, this does not hold for thermal transport properties like thermal conductivity and Nernst effect where charge carriers are always driven against the temperature gradient \cite{Parrott}. In addition, the observed sign change of $S_{yz}(H)$ for Mn$_5$Si$_3$C$_{0.8}$ rules out this explanation. An enhancement of the ANE and a sign change of the ANE coefficient was observed in the thickness dependence of thin ferromagnetic films \cite{Chuang}. While the enhancement was attributed to intrinsic and side-jump mechanisms the origin of the sign change remained unclear. Similarly, a sign change of $S_{yz}$ of unkown origin was reported to occur in the temperature dependence of Ga$_{1-x}$Mn$_{x}$As ferromagnetic semiconductors \cite{Pu}, as well as in pyrochlore molybdates \cite{Hanasaki2008} which has been attributed to a spin chirality contribution. In the present case, it is not clear why a sign change in the temperature dependence occurs only in two of the five films leaving this question unsolved.

\section{Discussion}
\begin{figure}
	\centering
	\includegraphics[width=\columnwidth]{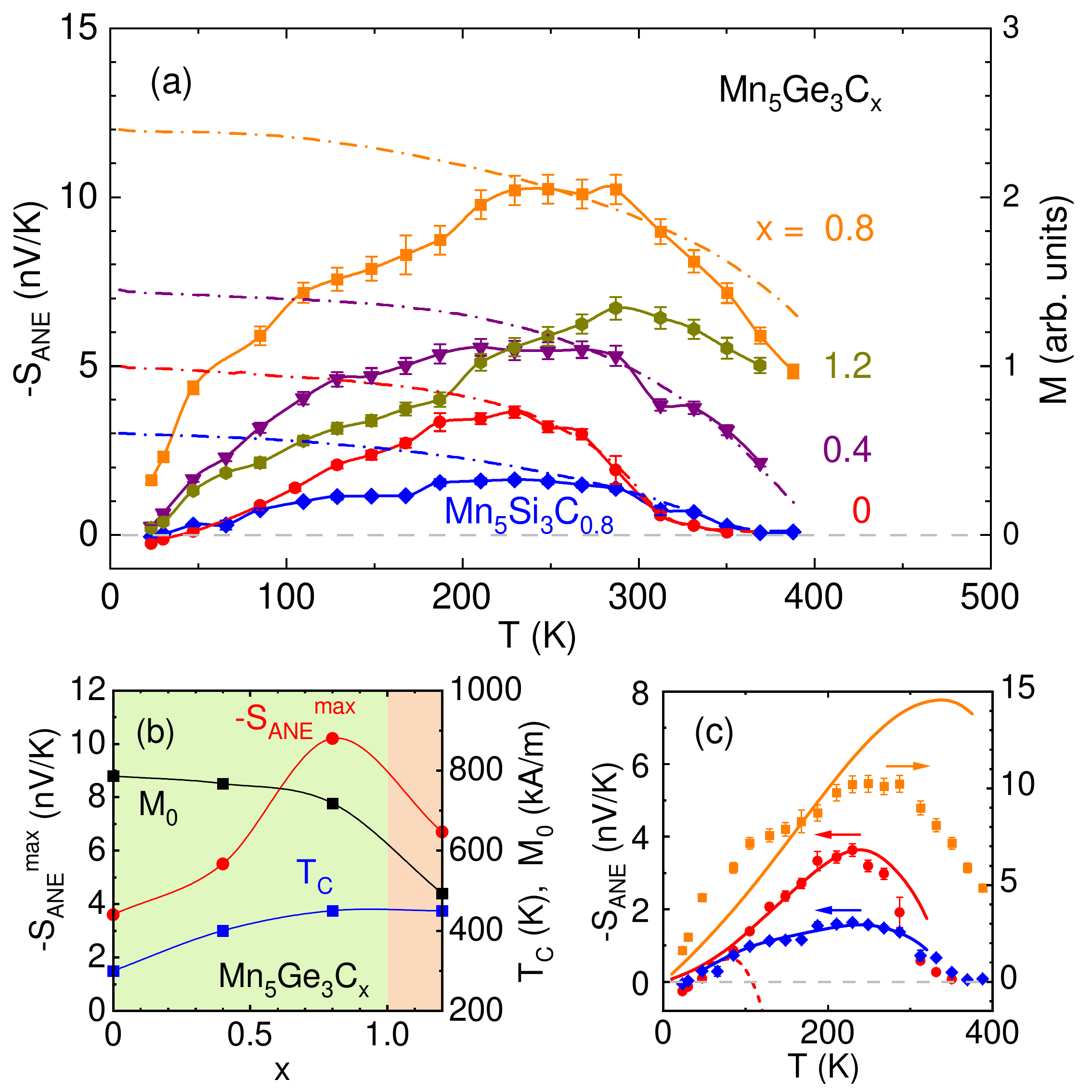}
	\caption{(a) ANE coefficient $S_{\rm ANE}$ vs.~temperature \textit{T}. Dash-dotted lines indicate the temperature dependence of the magnetization $M(T)$ for Mn$_5$Ge$_3$C$_{x}$ with $x=0$, $x=0.4$, and $x=0.8$, as well as Mn$_5$Si$_3$C$_{0.8}$. (b) Maximum ANE coefficient $S_{\rm ANE}^{\rm max}$, saturation magnetization $M_0$, and Curie temperature $T_{\rm C}$ vs.~carbon concentration $x$. (c) ANE coefficient. Dashed line shows the behavior $S_{\rm ANE}$ expected from theory, Eq.~\ref{ANEtheory}, solid line shows $S_{\rm ANE}$ without the contribution from the Seebeck coefficient $S_{zz}$.} 
\label{F4}
\end{figure}

The temperature dependence of the ANE coefficient $S_{\rm ANE}$, plotted in Fig.~\ref{F4}(a), shows a roughly linear behavior at low temperatures where according to Nernst's theorem, the thermopower and thus the Nernst effect has to vanish for $T\rightarrow 0$. $S_{\rm ANE}$ values have been taken from the measured hysteresis by extracting the mean voltage within the field range $0.4$ to $0.6\,$T, where saturation is observed in accordance with the magnetization measurements. A maximum $S_{\rm ANE}^{\rm max}$ is reached between $200\,$K and $300\,$K for all samples, followed by a decrease towards higher temperatures eventually reaching zero at $T_{\rm C}$ due to the strong decrease of the magnetization $M$, and hence the AHE and ANE, respectively, as seen by comparing the temperature dependences $S_{\rm ANE}(T)$ and $M(T)$ for Mn$_5$Ge$_3$ ($T_{\rm C}=300\,$K), Mn$_5$Ge$_3$C$_{0.4}$ ($T_{\rm C}\approx 400\,$K), and Mn$_5$Si$_3$C$_{0.8}$ ($T_{\rm C}=350\,$K). The strongest ANE is observed for Mn$_5$Ge$_3$C$_{0.8}$ with an almost three times enhanced $S_{\rm ANE}$ compared to the parent Mn$_5$Ge$_3$ compound. This enhancement cannot simply be explained by the higher $T_{\rm C}$ and thus demonstrates the extraordinary properties of this optimal carbon-doped compound in line with previous results on the magnetic properties of Mn$_5$Ge$_3$C$_{x}$ films. On the one hand, Fig.~\ref{F4}(b) clearly shows that the strongest $S_{\rm ANE}$ observed for $x=0.8$ is accompanied by reaching the maximum $T_{\rm C}=450\,$K with only a minor decrease of the saturation magnetization. On the other hand, the strong reduction of $S_{\rm ANE}$ and $M_0$ at $x=1.2$ is possibly due to the carbon-induced formation of different phases reducing the volume fraction of ferromagnetic Mn$_5$Ge$_3$C$_{x}$ and, hence, of both properties \cite{GajdzikJAP}. 

The $S_{\rm ANE}$ values in the nV/K range are rather small compared to data obtained on bulk samples, for instance Mn$_3$Sn, where a value of \SI{0.35}{\micro\volt\per\kelvin} at $300\,$K has been reported \cite{Ikhlas}. This is due to the huge difference in film thickness and, hence, thermal conductance between film and substrate \cite{UchidaJPCM}. For the present configuration with a temperature gradient perpendicular to the film most of the temperature-gradient distribution is generated across the substrate while only a tiny fraction of the heat difference drops across the thin film. For a quantitative analysis of $S_{\rm ANE}$ a measurement of the intrinsic temperature gradient across the film thickness is necessary \cite{Sola}. In the present case it is impossible to take into account the different thermal conductivities $\kappa$ and thermal resistances of the metal film, insulating substrate, interface(s), Kapton foil etc.~and their individual temperature dependences (partly unknown) in order to calculate intrinsic values of $S_{\rm ANE}$. However, in this work we focus on the qualitative behavior of the ANE and the effect of carbon doping on the size of $S_{\rm ANE}$ and therefore do not consider these corrections further. 

The origin of the ANE is discussed in close relation with the AHE, see \textit{e.g.}, Ref.~\onlinecite{Asamitsu} for a study of itinerant ferromagnets. Like the AHE, the ANE is also due to spin-orbit interaction and can be separated into an intrinsic contribution arising from the electronic band structure and the Berry phase curvature, a side jump and a skew scattering mechanism. For the films investigated here the AHE is dominated by the intrinsic effect and side-jump mechanism for which a behavior $\rho_{yx} \propto M\rho_{xx}^2$ is observed, as discussed above. Skew scattering would give rise to a behavior $\rho_{yx} \propto M\rho_{xx}$ \cite{Nagaosa}. 

The link between the electrical and thermoelectric transport coefficients is provided by the Mott relation, which is also valid for off-diagonal transport properties \cite{Pu}, $\alpha_{yz}$=$(\pi^2{\rm k}_{\rm B}^2/3{\rm e})T(\partial{\sigma_{yz}}/\partial{E})_{E_{\rm F}}$, where $\alpha_{yz} = \sigma_{zz}S_{yz}+\sigma_{yz}S_{zz}$ is the Peltier coefficient and $\sigma_{zz}$ and $\sigma_{yz}$ are the conductivity and the Hall conductivity, respectively, and $S_{zz}= E_z/{\nabla T_z}$ is the Seebeck coefficient. Since the anomalous part of the Hall effect can be expressed by a power-law scaling $\rho_{xy} = S_H M \rho_{xx}^n$ with $n \approx 2$, see Fig.~\ref{F2}(c) and Ref.~\onlinecite{Surgers}, the ANE coefficient is given by \cite{Pu} 

\begin{equation}\label{ANEtheory}
S_{yz}=\frac{\rho_{xy}}{\rho_{xx}}\left[\left(\frac{\pi^2{\rm k}_{\rm B}^2}{3{\rm e}}T\right) \frac{S_H'}{S_H}-(n-1)S_{zz}\right]\,.
\end{equation}

where the ratio between $S_H$ and its energy derivative $S_H'=(\partial S_H/\partial E)_{E_{\rm F}}$ serves as a free parameter. Apart from this ratio it is immediately clear that the temperature dependence of $S_{yz}$, shown in Fig.~\ref{F4}(a) and (c), is governed by the ratio $\rho_{yx}(T)/\rho_{xx}(T)$ which is increasing in going from Mn$_5$Si$_3$C$_{0.8}$ to Mn$_5$Ge$_3$ to Mn$_5$Ge$_3$C$_{0.8}$ (see Fig.~\ref{F2}(c). 

Fig.~\ref{F4}(c) shows that the calculated $S_{\rm ANE}$ values for Mn$_5$Ge$_3$ (dashed line) do not at all describe the measurements when we assume $S_{zz}$ = $S_{xx}$. $S_{xx}$ has been determined in a separate measurement by generating an in-plane thermal gradient along $x$ and measuring the thermovoltage along $x$. Rather, we obtain good agreement between the calculations and experimental data if the contribution from $S_{zz}$ in Eq.~\ref{ANEtheory} is \textit{not} taken into account. Although it is reasonable to assume isotropic coefficients $S_{zz}$ = $S_{xx}$ because the films are polycrystalline as proved by x-ray diffraction (not shown), this might not be valid for the present case. One reason could be a strong difference between the measured $S_{xx}$ in the plane of the film and the required $S_{zz}$ perpendicular to the film plane. Furthermore, for Mn$_5$Ge$_3$C$_{0.8}$ the calculation strongly differs from the measured data. It was not possible to obtain a better agreement by changing the ratio $S_H'/S_H$, in particular concerning the maximum of $S_{\rm ANE}(T)$ occurring at $\approx 250\,$K in contrast to the maximum around $350\,$K obtained in the calculation. We mention that experimental results from a second Mn$_5$Ge$_3$C$_{0.8}$ film and also a calculation performed with previous $\rho_{yx}$ and $\rho_{xx}$ data \cite{Surgers} exhibit the same discrepancy. A strong enhancement of the ANE of Mn$_5$Ge$_3$C$_{0.8}$ has also been observed for films deposited on semiconducting Ge(111) when compared to the parent Mn$_5$Ge$_3$. This is currently being investigated.

\section{Conclusion}
The ANE of ferromagnetic Mn$_5$Ge$_3$C$_{x}$ and Mn$_5$Si$_3$C$_{0.8}$ films on insulating sapphire has been studied. The ANE coefficient $S_{\rm ANE}$ of Mn$_5$Ge$_3$C$_{0.8}$ is strongly enhanced compared to Mn$_5$Ge$_3$C$_{x}$ films with $x \neq 0.8$. This enhancement is partly caused by the larger ratio of Hall resistivity and longitudinal resistivity due to the change of the electronic band structure and, hence, Berry phase curvature by carbon incorporation. The results are in line with earlier investigations of the magnetic properties of this material and further demonstrate that Mn$_5$Ge$_3$C$_{x}$ with $x=0.8$ sticks out as a material with improved magnetic properties when compared to the parent Mn$_5$Ge$_3$ compound. The carbon-induced enhancement of the ANE also observed for films deposited on semiconducting Ge(111) is under scrutiny.

\begin{acknowledgments}
The authors gratefully acknowledge financial support by the Deutsche Forschungsgemeinschaft (DFG) Project SU225/3-1.
\end{acknowledgments}

\section*{Data Availability}
The data that support the findings of this study are available within the article.

\end{document}